# The evolution of obscured AGN across cosmic time

*A large quasar demographic optical MOS survey targeting over 50 million AGN candidates*

Tanya Urrutia[1], Darshan Kakkad[2], Paula Sánchez-Sáez[3],
Mojtaba Raouf[4,5], Swayamtrupta Panda[6], Sarah E I Bosman[7], Francisco Pozo Nunez[8], Annagrazia Puglisi[9], Sophia Flury[10], Dragana Ilic[11], Andjelka B. Kovacevic[11], Mamta Pandey-Pommier[12], Giustina Vietri[13], Sarath Satheesh-Sheeba[14], Francesco Salvestrini[15], Susanna Bisogni[13], Eduardo Bañados[16], Ana Monreal Ibero[4], Sabine Thater[17], Pratika Dayal[18], Filippo D'Ammando[19], José Afonso[20], Paramita Barai[15], Valentin Ivanov[3]

[1] Leibniz-Institut für Astrophysik, Potsdam (AIP), An der Sternwarte 16, 14482, Potsdam, Germany
[2] Centre for Astrophysics Research, Department of Physics, Astronomy and Mathematics, University of Hertfordshire, Hatfield, AL10 9AB, UK
[3] European Southern Observatory, Karl-Schwarzschild-Str. 2, 85748 Garching bei München, Germany
[4] Leiden Observatory, Leiden University, P.O. Box 9513, 2300 RA Leiden, Netherlands
[5] Faculty of Aerospace Engineering, Delft University of Technology, Delft, The Netherlands
[6] International Gemini Observatory, NSF NOIRLab, Casilla 603, La Serena, Chile
[7] Institute for Theoretical Physics, University of Heidelberg, Philosophenweg 12, Heidelberg 69115, Germany
[8] Heidelberg Institute for Theoretical Studies, Astroinformatics, Schloss-Wolfsbrunnenweg 35, 69118 Heidelberg, Germany
[9] School of Physics and Astronomy, University of Southampton, Highfield, SO17 1BJ, UK
[10] Institute for Astronomy, Royal Observatory, University of Edinburgh, Edinburgh EH9 3HJ
[11] University of Belgrade, Faculty of Mathematics, Studentski trg 16, 11000 Belgrade, Serbia
[12] Pole Scientific, University Catholic of Lyon- University of Lyon, 10 place des Archives 69288, Lyon, France
[13] INAF IASF-Milano, Istituto di Astrofisica Spaziale e Fisica cosmica Milano, Via Alfonso Corti 12 20133
[14] Instituto de Astrofísica, Facultad de Ciencias Exactas, Universidad Andres Bello, Fernández Concha 700, 7591538 Las Condes, Santiago, Chile
[15] INAF Osservatorio Astronomico di Trieste, via G.B. Tiepolo 11, 34143 Trieste, Italy
[16] Max-Planck-Insititute für Astronomie, Königstuhl 17, 69117 Heidelberg, Germany
[17] Department of Astrophysics, University of Vienna, Türkenschanzstrasse 17, 1180 Vienna, Austria
[18] Canadian Institute for Theoretical Astrophysics, 60 St George St, University of Toronto, Toronto, ON M5S 3H8, Canada
[19] INAF-IRA Bologna, Via P. Gobetti 101, I-40129, Bologna, Italy
[20] Instituto de Astrofísica e Ciências do Espaço, Faculdade de Ciências, Universidade de Lisboa, Tapada da Ajuda, 1349-018 Lisboa, Portugal

# The evolution of obscured AGN across cosmic time

## 1. Introduction

An active galactic nucleus (AGN) represents a significant phase in the lifecycle of a galaxy. AGNs are pivotal not only for the growth of supermassive black holes (SMBHs) at their centers (Soltan 1982) but also in regulating star formation and the overall stellar mass of their host galaxies (di Matteo et al. 2005). The most luminous quasars are believed to be triggered by mergers (Sanders & Mirabel 1996; Urrutia et al. 2008). Such mergers can disrupt angular momentum conservation, allowing substantial amounts of gas to be funneled toward the galaxy's center to create a luminous AGN. Consequently, AGNs profoundly transform their host galaxies.

The spectral energy distribution (SED) of an AGN is typically blue, adhering to an accretion power law complemented by a UV thermal component from the accretion disk (Elvis 1994). Most optical selections of quasars exploit this characteristic by targeting blue color-color regions indicative of this SED behavior (Richards et al. 2002). However, quasars that are either reddened by dust or obscured by intervening material often evade detection through this color selection.

Typically, AGNs are thought to be shrouded by a toroidal structure located beyond the dust sublimation radius of the accretion disk (Urry & Padovani 1995). Whether we see AGNs as obscured or not solely depends on the viewing angle we have towards the central engine. While the toroidal material is likely clumpy and may be shaped by quasar winds (Elitzur et al. 2012; Hönig & Kishimoto 2017), this model has been statistically validated across various wavelength regimes. One intriguing consequence of this model is the concept of a "receding torus," which posits that the obscured fraction of AGNs diminishes as luminosity increases due to the sublimation radius extending outward in response to the radiation emitted by the quasar (Lawrence 1991, González-Martín et al. 2017)

The observational landscape is, however, more intricate. X-ray surveys indicate that locally, around two-thirds of AGNs are obscured, and this fraction decreases with rising luminosity, as anticipated (Lusso et al. 2012; Ricci et al. 2017). Paradoxically, some luminous, young AGNs driving outflows are often obscured yet exhibit weak X-ray emissions; they are often found in compact nuclei in Ultraluminous Infrared galaxies (ULIRGs, Garcia-Bernete et al. 2022). Even our nearest quasar, Mrk 231, falls into that category (Veilleux et al. 2013; Teng et al. 2014). Similarly, the fraction of lightly obscured quasars selected in the infrared increases with luminosity (Lacy et al. 2015). Depending on the selection methodology, a significant population of obscured quasars may be overlooked. Assessing their fraction is crucial, as most SMBH growth occurs within obscured systems (Martínez-Sansigre et al. 2005).

## 2. Open questions today

It is known that quasars, selected via different selection methods, display distinct AGN properties, even before addressing the question of obscuration. For example, infrared-selected quasars are often younger, exhibit higher star-formation rates, but cluster less than their radio-selected counterparts (Hickox et al. 2009). Multi-wavelength AGN demographic surveys are often employed to mitigate these selection effects. However, due to the inherent flux limit, achieving completeness across both redshift and the full obscuration fraction is challenging, except in the case of the deepest, high-resolution pencil-beam surveys, which, by necessity, cover a limited solid angle.

Nevertheless, a diverse array of obscured AGN classes has emerged beyond the typical Type 2 AGN: Type 2 QSOs (Norman et al. 2002), steep spectrum radio galaxies (Gu & Ai 2011), XBONGs (Kim et al. 2023), Hot DOGs (Assef et al. 2015), red quasars (Glikman et al. 2012, Fawcett et al. 2022) and FeLoBAL quasars (Farrah et al. 2010) among others. In particular, one of the most puzzling findings of the James Webb Space Telescope (JWST) has been the class named "Little Red Dots" (Matthee et al. 2024), some of which could also represent dust enshrouded early quasar growth (Torralba et al. 2025). In addition, the emerging special class of "Changing-Look AGN" (CL AGN), in which AGN change their type due to a change in accretion (Denney et al. 2014) has challenged traditional obscuration spectral classification schemes.

Large spectroscopic campaigns, including SDSS V (BHM; Kollmeier et al. 2025), WEAVE-LOFAR (Smith et al. 2016), DESI (Alexander et al. 2023), and 4MOST (Merloni et al. 2019), aim to study demographics across more than 1 million quasar candidates. Yet, whether limited by the depth of parent selection surveys (e.g., eROSITA limits of ~$5\times10^{-14}$ erg s$^{-1}$ cm$^{-2}$) or spectroscopic sensitivity, these programs remain incomplete at $z > 1$. As a result, they cannot resolve the evolution of AGN obscuration, particularly during the peak of accretion activity at Cosmic Noon. Even at low redshifts, we are still missing even the brightest obscured quasars, let alone achieving overall completeness, because they don't adhere to typical selection criteria. For example, a follow-up on Gaia stellar targets without proper motions revealed many red quasars hidden in the stellar locus (Heintz et al. 2020).

### 3. A large quasar demographics survey for the 2040s

*We propose a large quasar demographic optical multi-object spectroscopic (MOS) survey targeting over 50 million AGN candidates up to the highest redshifts possible in the optical (z~6.5), with repeat visits, using a variety of selection criteria available by 2040. A large MOS survey combining all AGN selection methods is the only way to unify a diverse range of different obscured AGN populations within a single, variability- and spectroscopy-based framework, rather than as disjoint classes selected by different methods.*

Space-based missions such as Roman and Euclid are set to image 2,000 and 14,000 square degrees, respectively, in the near-infrared. Given that Euclid's early data release already identified over 100,000 potential red quasars (Tarsitano et al. 2025), a high density of AGN targets is achievable with these missions alone. Furthermore, the SKA-Wide survey is expected to yield an additional 20–50 million quasar candidates, with over 50,000 extending into the Epoch of Reionization. Because radio surveys are largely insensitive to dust obscuration, they promise an unbiased view of the obscured quasar fraction across all luminosities and cosmic time. NewAthena will further enhance this landscape with deep X-ray surveys; these are crucial for high redshifts, where the telescope's energy range probes rest-frame hard X-rays. Finally, active black hole candidates identified through multi-messenger or astrometric methods, such as gravitational waves (LISA) or infrared astrometry (GaiaNIR), will complete the census.

In addition, Vera C. Rubin Observatory's Legacy Survey of Space and Time (LSST) will provide millions of candidates via intrinsic variability; a method less sensitive to obscuration. Dedicated spectroscopic monitoring of these candidates will uncover hundreds of thousands of CL AGN and probe the intrinsic accretion disk and Broad Line Region (BLR) in heavily reddened systems. This offers a new method to disentangle between orientation, evolutionary stage, and accretion state changes in obscured versus unobscured systems. Using multi-epoch spectroscopy to track obscuration changes, BLR re-emergence, and disk-wind signatures over time will provide unique, direct constraints on the physical origin and evolution of obscuration.

While identifying 50 million quasars yields an exceptionally large sample, such volume is essential to correct for selection biases. For instance, FeLoBALs were originally thought to be extremely rare based on the SDSS survey (>100,000 quasars), representing less than 1% of the population (Trump et al. 2006). However, radio and mid-infrared selection revealed the fraction to be significantly higher (~10%; Urrutia et al. 2009; Dai et al. 2012). To achieve completeness and ensure a statistically representative sample of at least 100 objects per class, luminosity, and redshift bin, a dataset of this magnitude is required.

## 4. Necessary technology developments

While current large MOS instruments capable of 1000–5000 simultaneous spectra are mounted on 2–4m class telescopes, the next generation of surveys demands significantly greater capability. Current facilities being developed for the ELT can observe only a relatively small number of targets in very deep fields. Characterizing over 50 million AGN targets, particularly at faint magnitudes and high redshifts, requires a facility combining high multiplexing (20,000–30,000 spectra) with a 10–15m aperture, such as the Wide-Field Spectroscopic Telescope (WST, Mainieri et al. 2024). This increased collecting area is critical for obtaining spectra of heavily obscured sources ($N_H \sim 10^{24}$ cm$^{-2}$), reaching down to 25th magnitude with low-medium spectral resolution (R~1500) and achieving a S/N > 5 in the potential broad emission and absorption lines across the entire optical wavelength regime.

Such a facility would come with its own inherent challenges, including a fast fiber positioner (de Jong et al, 2024) and the handling of the enormous data that such a large number of spectra presents, from readout and reduction to automatic identification and storage. Development in the 2030s should focus on overcoming these challenges toward a large spectroscopic facility that can also tackle other astrophysical subjects besides AGN.